\documentclass[12pt]{article}
\usepackage[a4paper, total={7in, 10in}]{geometry}
\usepackage[parfill]{parskip}
\usepackage{physics, tensor, float, subcaption}
\usepackage{graphicx}
\graphicspath{ {Plots/} }
\usepackage{jhep-mod}
\usepackage{bm}
\usepackage{soul}
\usepackage{amssymb,amsmath,amsthm}
\usepackage{mathrsfs}
\usepackage[utf8]{inputenc}
\usepackage{enumerate}
\usepackage{bigints}
\usepackage{xcolor}
\usepackage{appendix}
\usepackage{graphicx}
\usepackage{float}
\usepackage{tikz}
\usepackage{setspace}
\usepackage{cancel}
\usepackage{doi}
\definecolor{purple}{rgb}{1,0,1}
\definecolor{lime}{HTML}{A6CE39} 

\newcommand{\blue}[1]{{\color{blue} #1}}

\definecolor{lime}{HTML}{A6CE39}
\newcommand{\orcidicon}{%
	\begin{tikzpicture}
	\draw[lime, fill=lime] (0,0) 
		circle [radius=0.16] 
		node[white] {{\fontfamily{qag}\selectfont \tiny ID}};
	\draw[white, fill=white] (-0.0625,0.095) 
		circle [radius=0.007];
	\end{tikzpicture}
	\hspace{-5mm}
}

\newcommand\orcidMatt{{\href{https://orcid.org/0000-0003-1088-6485}{\orcidicon}}}

\newcommand{\e}{\mathrm{e}}

\renewcommand{\O}{\mathcal{O}}
\newcommand\D{{\cal D}} 
\begin{document}
\title{
Feynman's $i\epsilon$ prescription,\\
almost real spacetimes,\\
and acceptable complex spacetimes
}
\author{
\Large
Matt Visser\!\orcidMatt\!
}
\affiliation{School of Mathematics and Statistics, Victoria University of Wellington, \\
\null\qquad PO Box 600, Wellington 6140, New Zealand.}
\emailAdd{matt.visser@sms.vuw.ac.nz}

\abstract{
\vspace{1em}

Feynman's $i\epsilon$ prescription for quantum field theoretic propagators has a quite natural reinterpretation in terms of a slight complex deformation of the Minkowski spacetime metric. 
Though originally a strictly flat-space result, once reinterpreted in this way, these ideas can be naturally extended first to semi-classical curved-spacetime  QFT on a fixed background geometry and then, (with more work),
to fluctuating spacetime geometries. 
There are intimate connections with variants of the weak energy condition. We shall take the Lorentzian signature metric as primary, but note that allowing the complex deformation to become large leads to a variant of Wick rotation, and more importantly leads to physically motivated  constraints on the configuration space of acceptable off-shell geometries to include in Feynman's functional integral when attempting to quantize gravity. 
Ultimately this observation allows one to connect the discussion back to recent ideas on ``acceptable'' complex metrics, in the Louko--Sorkin and  Kontsevich--Segal--Witten sense, with Lorentzian signature spacetimes occurring exactly on the boundary of the set of ``acceptable'' complex metrics. By adopting the tetrad formalism we explicitly construct the most general set of acceptable complex metrics satisfying the 0-form, 1-form, and 2-form acceptability conditions. \bigskip

\bigskip
\noindent
{\sc Date:} 28 November 2021; 5 December 2021;
9 June 2022; 13 July 2022;\\
\LaTeX-ed: \today \ (Final version accepted for publication in JHEP.)

\bigskip
\noindent{\sc Keywords}: Complex metrics; acceptability conditions; Feynman $i\epsilon$ prescription; Feynman propagator; Wick rotation; Lorentzian signature; Euclidean signature; tetrad formalism.

\bigskip
\noindent{\sc PhySH:} 
Gravitation
}

\maketitle
\def\tr{{\mathrm{tr}}}
\def\diag{{\mathrm{diag}}}
\def\cof{{\mathrm{cof}}}
\def\pdet{{\mathrm{pdet}}}
\def\d{{\mathrm{d}}}
\def\L{{\mathcal{L}}}
\def\S{{\mathcal{S}}}
\def\Z{{\mathcal{Z}}}
\def\Wick{{\mathrm{Wick}}}
\def\scalar{{\mathrm{scalar}}}
\def\gauge{{\mathrm{gauge}}}
\def\Proca{{\mathrm{Proca}}}
\parindent0pt
\parskip7pt
\def\Kerr{{\scriptscriptstyle{\mathrm{Kerr}}}}
\def\eos{{\scriptscriptstyle{\mathrm{eos}}}}
\def\e{{\mathrm{e}}}

\clearpage
\section{Introduction}

Feynman's $i\epsilon$ prescription for the Feynman propagator was originally developed as a pragmatic trick for encoding causality into the Fourier transformed momentum-space propagators occurring in the Feynman diagram expansion~\cite{Bjorken-Drell,Lifshitz,Itzykson-Zuber,Peskin,Srednicki}. Feynman's $i\epsilon$ prescription then justifies flat-space Wick rotation, and is an essential ingredient in Euclideanizing any loop integrals that might be of interest~\cite{Bjorken-Drell,Lifshitz,Itzykson-Zuber,Peskin,Srednicki}. Going beyond the Feynman diagram expansion there has now been an almost 50 year history of reinterpreting Feynman's $i\epsilon$ prescription  in position space, often via a slight complex deformation of the Minkowski metric~\cite{Candelas:1977, Ivashchuk:1987, Ivashchuk:1988, Ivashchuk:1987-2019, Ivashchuk:1988-2002, Ivashchuk:1997,Visser:wick}. 

We shall soon see that in flat Minkowski spacetime, using  $(-+++)$ signature, for small $\epsilon$ it is useful to focus on the almost-real metric
\begin{equation}
(\eta_\epsilon)_{ab} = \eta_{ab} +i\epsilon\,V_a\,V_b 
+\O(\epsilon^2).
\end{equation}
Equivalently, for the inverse metric
\begin{equation}
(\eta_\epsilon)^{ab} = \eta^{ab} -i\epsilon\,V^a\,V^b
+\O(\epsilon^2).
\end{equation}
Here $V^a$ is at this stage some arbitrary but fixed spatially constant 4-velocity, while $\epsilon$ is a small position-dependent constant. 

Viewed as a slight complex deformation of the Minkowski metric, formal convergence of the Feynman functional integral~\cite{Glimm-Jaffe} can then be related to the whether or not the underlying classical Lagrangian appearing in the path integral satisfies an integrated off-shell variant of the weak energy condition (WEC). 

It is then tolerably straightforward to extend these ideas to curved spacetimes. First to curved-spacetime QFT on a fixed background, 
and then with considerably more work to fluctuating spacetime geometries.
In curved space for small $\epsilon$ it is useful to focus on the almost-real metric
\begin{equation}
(g_\epsilon)_{ab} = (g_L)_{ab} +i\epsilon\,V_a\,V_b +\O(\epsilon^2).
\end{equation}
Equivalently, for the inverse metric
\begin{equation}
(g_\epsilon)^{ab} = (g_L)^{ab} -i\epsilon\,V^a\,V^b +\O(\epsilon^2).
\end{equation}
Here $(g_L)_{ab} $ is now  some arbitrary Lorentzian signature metric, and  $V^a$ is now some arbitrary field of unit-norm 4-velocities.
Furthermore $\epsilon(x)>0$ is now allowed to be position-dependent.
We shall take the existence of the Lorentzian signature metric $(g_L)_{ab}$ as primary, and view the complex metric $(g_\epsilon)_{ab}$ as auxiliary.

\clearpage
We shall then further generalize this construction,  to considering ``allowable'' complex metrics of the form
\begin{equation}
\sqrt{-g_\epsilon}\, (g_\epsilon)^{ab} = 
\sqrt{-g_L} \,(g_L)^{ab} - i \epsilon \sqrt{g_E}\, (g_E)^{ab}.
\end{equation}
Here $(g_L)^{ab}$ and $(g_E)^{ab}$ are Lorentzian signature and Euclidean signature metrics respectively, and are subject to the additional nonlinear acceptability condition \begin{equation}
\Im\left\{\sqrt{-\det\left[ \sqrt{-g_L} \,(g_L)^{ab} 
- i \epsilon \sqrt{g_E}\, (g_E)^{ab}\right]} \right\} <0.
\end{equation}
These two conditions are a re-phrasing and explicit codification of the constraints first discussed by Louko and Sorkin~\cite{Louko:1995}, (the 0-form and 1-form constraints), and then recently revived  (and extended) by Kontsevich and Segal~\cite{Segal:2021}, and Witten~\cite{Witten:2021}. 
When presented in this way, the fact that Lorentzian signature metrics lie on the boundary of the acceptable complex metrics is explicitly manifest. 
Though the above is the most symmetric way of phrasing things, it  may not always be the most computationally effective, and we shall put some effort into various simplifications. 

For instance, let $h^{ab}$ be an arbitrary positive definite tensor, then one can recast the two Louko--Sorkin acceptability conditions in a somewhat  asymmetric but more computationally efficient manner: 
\begin{equation}
\label{E:master-intro}
 (g_\epsilon)^{ab} = {
 (g_L)^{ab} - i \epsilon \; h^{ab}
 \over
 \sqrt{\det\left\{
\delta_c{}^e -  i \epsilon \; (g_L)_{cd}h^{de} 
\right\}
}
}.
\end{equation}
\begin{equation}
\label{E:LS2-intro}
\Im \left( \sqrt{\det\left\{
\delta_a{}^c -  i \epsilon \;  (g_L)_{ab} h^{bc} 
\right\}}
\right)
< 0.
\end{equation}

\enlargethispage{40pt}
We also discuss the small-$\epsilon$ version of the above. 
For positive definite $h^{ab}$: 
\begin{equation}
(g_\epsilon)^{ab} = 
 (g_L)^{ab} - i \epsilon \; \overline{h^{ab}}
+ \O(\epsilon^2);
\qquad\qquad
(g_L)_{cd} \; h^{cd} > \O(\epsilon);
\end{equation}
where $\overline{h^{ab}}$ denotes the usual trace-reversal process.

We shall then, following and extending the ideas of Louko and Sorkin~\cite{Louko:1995}, and Kontsevich and Segal~\cite{Segal:2021}, and Witten~\cite{Witten:2021}, discuss the third (2-form) acceptability constraint relevant in a 4-dimensional setting, and use it to fully characterize the positive definite tensor $h^{ab}$ by constraining its eigenvalues. (In higher $D$-dimensional spacetime one would need to consider additional constraints, up to a $[D/2]$-form constraint.)
Then, adopting the tetrad formalism, we shall provide a complete and explicit characterization of all complex metrics satisfying the compatibility conditions. 

Finally we make some general comments about the generic functional integration over complex manifolds, and in the appendix present a fully explicit example of a complex spacetime metric that is \emph{not}  acceptable. 

\clearpage
\section{Flat space-time sign conventions}

Maintaining consistent sign conventions is somewhat tedious. 
Sign conventions are not entirely uniform even within the high-energy and/or general relativity communities, let alone between those communities.
I shall attempt to maximize coherence with major textbook references. 
While keeping track of overall sign conventions is annoying, this issue really does not greatly impact the ultimate physics conclusions. 
Let us first consider flat Minkowski spacetime, subsequent generalizations, (to curved spacetime and complex metrics, the bulk of the article), will (at least in principle) be straightforward.

\subsection{Metric}
We shall choose our metric conventions such that the flat
space Minkowski metric is $\eta_L = {\rm diag}(-1,+1,+1,+1)$. The flat space
Euclidean metric is taken to be positive definite, so that we have $\eta_E = {\rm
diag}(+1,+1,+1,+1)$. Particle physicists often use the opposite signature $\eta_L = {\rm diag}(+1,-1,-1,-1)$, which can sometimes lead to annoying stray minus signs.
For instance, Louko and Sorkin~\cite{Louko:1995} use 
$(-+++)$ signature whereas Kontsevich and Segal~\cite{Segal:2021}, and Witten~\cite{Witten:2021} use $(+---)$ signature. 
To add to the potential confusion, some relativists also use $(+---)$ signature; see the flyleaf of Misner--Thorne--Wheeler~\cite{MTW}  for some examples. 
Thence the phrases ``east coast metric'' and ``west coast metric''.

\subsection{Scalar Lagrangian}

There are also some tricky sign conventions to keep in mind as regards the relevant scalar Lagrangians $\L$. 
\begin{itemize}
\item 
For the Lorentzian signature scalar field, regardless of whether one is using $(-+++)$ or $(+---)$ signature,  one should always choose 
\begin{equation}
\L_L = {1\over 2} (\dot \phi )^2 - {1\over2}\sum_i  (\partial_i \phi)^2 - V(\phi).
\end{equation}
This will guarantee positive-semidefinite gradient contributions to the energy density
\begin{equation}
\rho_L = +{1\over 2} (\dot \phi )^2 + {1\over2}\sum_i  (\partial_i \phi)^2 + V(\phi) 
= + {1\over2}\sum_a  (\partial_a \phi)^2 + V(\phi).
\end{equation}

\clearpage
Thence a naive Wick rotation for the energy $E\to +i E$ implies $t\to -it$ which in turn  yields $(\dot\phi)^2 \to - (\dot\phi)^2$, and so
\begin{equation}
\L_\Wick = -{1\over 2} (\dot \phi )^2 - {1\over2}\sum_i  (\partial_i \phi)^2 - V(\phi) 
= - {1\over2}\sum_a  (\partial_a \phi)^2 - V(\phi).
\end{equation}
(At this preliminary stage of the discussion one can even be safely ambiguous as to the sign of the naive Wick rotation $t \to \pm it$, since then $\dot\phi \to \mp i \dot \phi$, and so in either case $(\dot\phi)^2 \to - (\dot\phi)^2$. We shall subsequently need to be more careful.)
The fact that $\L_\Wick$ has negative-semidefinite gradient contributions is inconvenient, typically one flips the overall sign and simply \emph{defines} the Euclidean scalar Lagrangian as
\begin{equation}
\L_E = - \L_\Wick = +{1\over 2} (\dot \phi )^2 + {1\over2}\sum_i  (\partial_i \phi)^2 + V(\phi) 
= + {1\over2}\sum_a  (\partial_a \phi)^2 + V(\phi).
\end{equation}

\item
Similarly for a gauge field (either Abelian or non-Abelian), 
regardless of whether one is using $(-+++)$ or $(+---)$ signature, one has
\begin{equation}
\L_L = {1\over2} \{ E^2-B^2\}; \qquad 
\rho_L = {1\over2} \{ E^2+B^2\}.
\end{equation}
Thence a naive Wick rotation of the energy $E\to i E$ implies $t\to -it$ which in turn for the Electric field yields $E\to - i E$ so that
\begin{equation}
\L_\Wick = -{1\over 2} \{ E^2+B^2\}.
\end{equation}
(It is easiest to carry this out in temporal gauge where $A_0=0$; in general for a gauge field Wick rotation corresponds to $t\to -it$ combined with $A_0 \to +i A_0$.
At this preliminary stage of the discussion one can even be safely ambiguous as to the sign of the naive Wick rotation $t \to \pm it$, $A_0\to \mp i A_0$, since then, using $E_i= \partial_i A_0-\dot A_i $, we have  $E \to \mp i E$, and so in either case $E^2 \to - E^2$. We shall subsequently need to be more careful.)
The fact that $\L_\Wick$ is negative-semidefinite is inconvenient, typically one flips the overall sign and \emph{defines} the Euclidean gauge Lagrangian as
\begin{equation}
\L_E = - \L_\Wick = +{1\over 2}\{ E^2+B^2\}.
\end{equation}

\clearpage
\item
Finally, if one now wishes to add a Proca mass term to the (Abelian) gauge Lagrangian then, regardless of whether one is using $(-+++)$ or $(+---)$ signature, one has the extra contribution
\begin{equation}
\L_L = -{1\over2} m_\gamma^2 \left\{ -A_0^2 + \sum_i A_i^2 \right\};
\qquad
\rho_L = {1\over2} m_\gamma^2 \left\{ -A_0^2 + \sum_i A_i^2 \right\}.
\end{equation}
This choice implies that the usual transverse polarization modes make a manifestly positive contribution to the energy density. (Checking positive energy density for the third longitudinal mode is straightforward.) 
Thence a naive Wick rotation, now of the form $A_0 \to +i A_0$, yields
\begin{equation}
\L_\Wick = -{1\over 2} m_\gamma^2 \left\{ +A_0^2 + \sum_i A_i^2\right\}= 
-{1\over 2} \left\{ \sum_a A_a^2\right\}
\end{equation}
(At this preliminary stage of the discussion one can even be safely ambiguous as to the sign of the naive Wick rotation $A_0\to \mp i A_0$, since in either case $(A_0)^2 \to - (A_0)^2$. We shall subsequently need to be more careful.)
The fact that $\L_\Wick$ is negative-semidefinite is inconvenient, typically one flips the sign and defines a possible Proca term contribution to the Euclidean gauge Lagrangian as
\begin{equation}
\L_E = - \L_\Wick = +{1\over 2} m_\gamma^2 \left\{ \sum_a A_a^2\right\}.
\end{equation}

\end{itemize}

By presenting the discussion in this manner we have been able to completely side-step the question of whether one is using $(-+++)$ or $(+---)$ signature; and have even (so far) been able to side-step the question of the overall sign of the Wick rotation $t \to \pm it$. Essentially all relevant sources will agree with the discussion above.

\subsection{Volume element}

One now needs to think about the volume element $\sqrt{-\det{\eta_{ab}}}\; \d^4 x$ and its Euclidean relative $\sqrt{\det{\delta_{ab} }}\; \d^4 x$. Though not normally phrased this way one has to decide whether to view the naive Wick rotation as a passive or active transformation.
\begin{description}
\item[Passive:] 
If you view the naive Wick rotation $t\to\pm it$ as \emph{passive} then one has both $\sqrt{-\det{\eta_{ab}}}\to \mp i \sqrt{\det{\delta_{ab} }}$ and $\d^4 x \to \pm i \d^4 x$. 
Then one would have $\sqrt{-\det{\eta_{ab}}}\;\d^4 x\to  \sqrt{\det{\delta_{ab} }}\;\d^4 x$, which proves  uninteresting.
\item[Active:]
If you view the naive Wick rotation $t\to\pm it$ as \emph{active} then one has  $\sqrt{-\det{\eta_{ab}}}\to \pm i \sqrt{\det{\delta_{ab} }}$, whereas $\d^4 x \to \d^4 x$ is unaffected. 
Then one would have $\sqrt{-\det{\eta_{ab}}}\;\d^4 x\to  \pm  i\sqrt{\det{\delta_{ab} }}\;\d^4 x$, which now does lead to interesting behaviour for the action.
\end{description}

\clearpage
\subsection{Action}

With the conventions above, in Cartesian coordinates one always has
\begin{equation}
\S_L = \int \L_L\; \sqrt{-\det[\eta_L]}\,\d^4 x; \qquad 
\S_E = \int\L_E\; \sqrt{\det[\delta]} \,\d^4 x \;. 
\end{equation}
Under a naive (active) Wick transformation $t\to\pm it$ one then has
\begin{eqnarray}
\S_L &=& \int \L_L\; \sqrt{-\det[\eta_L]}\,\d^4 x\to 
\int (- \L_\Wick)\; (\pm i\sqrt{\det[\delta_{ab}]})\,\d^4 x
\nonumber\\
&=&
\mp i \int \L_E\; \sqrt{\det[\delta_{ab}]}\,\d^4 x
= \mp i \S_E
\end{eqnarray}
This observation will ultimately allow us to fix the \emph{sign} in the naive Wick transformation $t\to\pm it$.

In curvilinear coordinates (but still in flat Minkowski space) this becomes
\begin{equation}
\S_L = \int \L_L\;  \sqrt{-g_L} \, \d^4 x; \qquad 
\S_E = \int \L_E \sqrt{g_E} \,\d^4 x, 
\end{equation}
where $\S_L\to\mp i \S_E$. 
Essentially all relevant sources will agree with these formulae.

\enlargethispage{20pt}
\subsection{Path integral}

With the conventions presented above above, the Feynman functional integral for the (flat-spacetime) partition function becomes
\begin{equation}
\Z_L = \int \e^{i S_L} \; \D\hbox{(fields)};
\qquad\qquad
\Z_E = \int \e^{- S_E} \; \D\hbox{(fields)}.
\end{equation}
This now makes it clear that we need to choose the $-$ sign in the naive Wick transformation $t\to-it$. (Equivalently, in terms of energy, $E\to +i E$.)  
On the other hand, here $\S_L$ and $\S_E$ have been set up in such a manner that we have been able to completely side-step (so far) the question of whether one is using $(-+++)$ or $(+---)$ signature.
Essentially all relevant sources will agree with these formulae.

\subsection{Curved spacetimes}
Finally in (real-metric) curved spacetime one (in principle) just replaces $\eta_{ab} \to (g_L)_{ab}$, or replaces $\delta_{ab} \to (g_E)_{ab}$, as appropriate.
However the ultimate goal of the current article is to reinterpret the naive Wick transformation in terms of a complex metric and to thereby 
fully characterize the complete set of all physically acceptable \emph{complex} metrics, a task to which we shall now turn. 

\clearpage
\section{Complexification of flat Minkowski spacetime}
\subsection{Feynman propagator: Small $\epsilon$}

The Feynman propagator in momentum space (for simplicity let us consider a massive scalar field) can be written as~\cite{Bjorken-Drell,Lifshitz,Itzykson-Zuber,Peskin,Srednicki}~\footnote {$E$ now denotes energy rather than electric field; the meaning should be clear from context.}
\begin{equation}
\label{E:P1}
\Delta_F(E,p) =
{i\over E^2-p^2-m^2+i\epsilon}.
\end{equation}
The whole point of Feynman's $i\epsilon$ prescription is to ``dodge'' around the poles, naively located at $E=\pm \sqrt{m^2+p^2}$, in an appropriate manner, by shifting the poles slightly to $E = \pm\sqrt{m^2+p^2-i\epsilon}$, see for instance~\cite{Bjorken-Drell,Lifshitz,Itzykson-Zuber,Peskin,Srednicki}.
\smallskip

Since both of these poles occur at $E^2 >0$, one could just as easily write
\begin{equation}
\label{E:P2}
\Delta_F(E,p) =
{i\over E^2(1+i\epsilon)-p^2-m^2}.
\end{equation}
When the propagator is written in
this form the $i\epsilon$ prescription has a natural interpretation in terms
of a complex ``not quite Minkowski'' metric~\cite{Ivashchuk:1987, Ivashchuk:1988, Ivashchuk:1987-2019, Ivashchuk:1988-2002, Ivashchuk:1997,Visser:wick}.  Specifically, let us define the slightly complex metric\footnote{We have now made the specific signature choice that as $\epsilon\to0$ the signature becomes $(-+++)$.}
\begin{equation}
(\eta_\epsilon)^{ab} = {\rm diag} (-1-i\epsilon, +1, +1, +1),
\end{equation}
and consider the 4-momentum covector $P_a =(-E,p_i)$.
That the 4-momentum covector is the appropriate quantity to work with is ultimately due to the fact that the kinetic operator in the Lagrangian is $\eta^{ab} \partial_a \partial_b$. 
Indeed, the quantum propagator (\ref{E:P1})\break
 ultimately arises from Fourier transforming the quantum fields. Thence using the standard plane-wave conventions $\exp( -i[\omega t - \vec k\cdot \vec x])$ focusses one's attention on the covariant wave 4-vector $K_a = (-\omega; \vec k)$. Then one has  $K^a =(+\omega; \vec k)$, and for the (virtual) 4-momentum $P^a = \hbar\, K^a = (E;\vec p)$. 

Thence the Feynman propagator can be re-written as
\begin{equation}
\label{E:P3}
\Delta_F(E,p) =
{i\over -(\eta_\epsilon)^{ab}  P_a P_b-m^2}.
\end{equation}

These three propagators, (\ref{E:P1})--(\ref{E:P2})--(\ref{E:P3}), all carry and imply exactly the same physics content~\cite{Ivashchuk:1987, Ivashchuk:1988, 
Ivashchuk:1987-2019, Ivashchuk:1988-2002, Ivashchuk:1997, Visser:wick}. Defining the 4-velocity $V^a =(1,0,0,0)$ we can write
\begin{equation}
(\eta_\epsilon)^{ab} = \eta^{ab} -i\epsilon\,V^a\,V^b.
\end{equation}
This version of Feynman's $i\epsilon$ prescription, where the $\epsilon$ has been pushed into a slight complexification of the spacetime metric, has a much better chance of being usefully generalizable. Since at this stage we are assuming $\epsilon$ to be arbitrarily small we could just as well write
\begin{equation}
(\eta_\epsilon)^{ab} = \eta^{ab} -i\epsilon\,V^a\,V^b +\O(\epsilon^2).
\end{equation}
Equivalently, for the covariant metric one takes the matrix inverse
\begin{equation}
(\eta_\epsilon)_{ab} = \eta_{ab} +i\epsilon\,V_a\,V_b +\O(\epsilon^2).
\end{equation}

For future use we note the two determinants~\footnote{These two determinants are independent of whether one chooses $(-+++)$ or $(+---)$ signature.}
\begin{equation}
\det[(\eta_\epsilon)^{ab}] = -1-i \epsilon +\O(\epsilon^2),
\end{equation}
and the more directly useful
\begin{equation}
\det[(\eta_\epsilon)_{ab}] = -1+i \epsilon +\O(\epsilon^2),
\end{equation}
so that
\begin{equation}
\sqrt{-\det[(\eta_\epsilon)_{ab}]} = 1-{1\over2} i \epsilon +\O(\epsilon^2).
\end{equation}
This is explicitly compatible with the footnote on page 193 of the Louko--Sorkin~\cite{Louko:1995} article.

Consequently
\begin{eqnarray}
\sqrt{-\det[(\eta_\epsilon)_{ab}]} \;\; (\eta_\epsilon)^{ab} 
&=& 
 \left(1-{1\over2} i \epsilon + \O(\epsilon^2) \right)\; 
\left( \eta^{ab} -i\epsilon\,V^a\,V^b +\O(\epsilon^2) \right)
\nonumber\\
&=& \eta^{ab} -i\epsilon\,\left[V^a\,V^b+{1\over2} \eta^{ab}\right] +\O(\epsilon^2).
\end{eqnarray}
Here in $(-+++)$ signature the quantity $\left[V^a\,V^b+{1\over2} \eta^{ab}\right] $ is guaranteed to be a positive definite matrix. 
This is explicitly compatible with the footnote on page 193 of the Louko--Sorkin~\cite{Louko:1995} article.
This small-$\epsilon$ flat-space result is the key observation that we shall seek to generalize, first to large $\epsilon$ and then to curved spacetimes.

\clearpage
\subsection{Flat-space Wick rotation: Large $\epsilon$}

While this is not the central point of the current article, there are much deeper connections between this slightly complex metric and the idea of Wick rotation.
When performing the usual form of flat-space
Wick rotation, $E\to iE$, the contour does not pass over the poles. [The
contour $(-\infty,+\infty)$ is deformed to $(-i\infty,+i\infty)$.
So the contour dodges above the positive energy pole and below the negative energy pole.]  In terms
of the ``rotated'' energy variable, once one takes the limit $\epsilon\to0$, the Euclidean propagator is simply:
\begin{equation}
\Delta_E(E,p) = {-i\over E^2 +p^2 +m^2}\;.
\end{equation} 

In counterpoint, starting from an almost-real  complex metric, the energy $E$ instead remains untouched and for the  (contravariant) metric one has $\eta_\epsilon=\eta-i\epsilon V\otimes V+\O(\epsilon^2)$, that is $[\eta_\epsilon]^{ab} =\eta^{ab} -i\epsilon V^a V^b+\O(\epsilon^2)$, for $\epsilon$ small and positive.
To extend this to a ``large $\epsilon$'' version of the complex metric one would instead write $\eta_\epsilon=\eta-i\epsilon V\otimes V$, that is $[\eta_\epsilon]^{ab} =\eta^{ab} -i\epsilon V^a V^b$, 
 and would allow $\epsilon$ to become large and complex while
 keeping the real part of $\epsilon$ non-negative. Specifically one would
let $\epsilon$ travel from 0 to $2i$ while, (thanks to he real part of $\epsilon$ being non-negative), carefully dodging around the point $\epsilon=i$ where the complex metric would become singular.

Alternatively one could instead write an explicit ``large $\epsilon$'' version of  the complex metric as follows:
\begin{equation}
(\eta_\epsilon)^{ab} = \eta^{ab} +\{1-\exp(i\epsilon)\}\,V^a\,V^b,
\end{equation}
and let $\epsilon$ travel from 0 to $\pi$ along the positive real axis. 
This version of the complexified Minkowski metric is closer in spirit to the construction adopted in a series of papers by Greensite and Carlini~\cite{Greensite:1992, Greensite:1993, Carlini:1993, Greensite:1994, Greensite:1995}.

Another nice feature of putting the $i\epsilon$ into the spacetime metric is that it will then automatically take care of the polarization factors for higher spin. (Otherwise one would have to add somewhat \emph{ad hoc} modifications to Wick rotation by, for instance, additionally Wick rotating the electromagnetic scalar potential, $A^0\to -i A^0$ but not the vector potential $\vec A$, \emph{etc.})

For this ``large $\epsilon$'' version of  the complex metric 
one has
\begin{equation}
\det[ (\eta_\epsilon)^{ab}] = - \e^{i\epsilon}; \qquad
\det[ (\eta_\epsilon)_{ab}] = - \e^{-i\epsilon}; \qquad
\end{equation}
and consequently
\begin{eqnarray}
\sqrt{-\det[(\eta_\epsilon)_{ab}]} \;\; (\eta_\epsilon)^{ab} 
&=& 
 \e^{-i\epsilon/2}\; 
\left( \eta^{ab} +\{1-\e^{i\epsilon}\}\,V^a\,V^b \right)
\nonumber\\
&=& 
- \e^{i\epsilon/2} V^a V^b 
+\e^{-i\epsilon/2} [\eta^{ab} + V^a V^b]. 
\end{eqnarray}
We shall ultimately extend this ``large $\epsilon$'' version of  the complex metric to curved spacetimes.

\subsection{Weak energy condition}

Within the context of the Feynman functional integral~\cite{Glimm-Jaffe}, still working in flat Minkowski space,  the Lorentzian partition function is
\begin{equation}
Z_L = \int \D \phi \; \exp\left( + i S(\phi,\eta) \right).    
\end{equation}

Now $i\epsilon$-complexify the metric
\begin{equation}
Z_\epsilon = \int \D \phi \; \exp\left( + i S(\phi,\eta_\epsilon) \right).    
\end{equation}
For small $\epsilon$ we now use the general result that in $(-+++)$ signature
\begin{equation}
{\delta S(g,\phi)\over \delta g_{ab}} = - {\sqrt{-g}\over 2} \;\; T(g,\phi)^{ab},
\end{equation}
to write
\begin{equation}
S(\eta_\epsilon,\phi) = S(\eta,\phi) 
+ {i\epsilon\over2}  \int T^{ab}(\eta,\phi) \;V_a \,V_b \;\sqrt{-\eta} \;   d^4 x 
+\O(\epsilon^2).
\end{equation}
This now implies
\begin{equation}
Z_\epsilon = \int \D \phi \left\{ 
\exp\left(+ i S(\phi,\eta_\epsilon) \right)\;
\exp\left( - {\epsilon\over2}  \int T^{ab}(\eta,\phi) \;V_a \,V_b \;\sqrt{-\eta} \;   d^4 x 
+\O(\epsilon^2) \right)
\right\}.    
\end{equation}
Thence, if the classical off-shell stress-energy tensor satisfies the WEC, (so that $T^{ab}(\eta,\phi) \;V_a \,V_b \geq 0$), then the $i\epsilon$-complexified metric serves to damp the oscillations in the functional integral, which is exactly what we want to make the functional integral mathematically justifiable~\cite{Glimm-Jaffe}. (For related comments see also references~\cite{Louko:1995} and~\cite{Kothawala:2018}.)

Unfortunately the argument cannot be run in reverse. Even with a damped functional integral the expectation value $\langle T^{ab}(\eta,\phi) \rangle$ need not satisfy the WEC, both because of the phases $\exp\left( -i S(\phi,\eta_\epsilon) \right)$, and because of the need to renormalize.\footnote{And if you try to Wick rotate \emph{first}, there are other issues to deal with. Yes, the phases are now all real, but once one is in Euclidean signature the spacelike/timelike/null distinction no longer applies, and NEC/WEC/SEC/DEC are either ill-defined or all collapse to the SEC~\cite{Martin-Moruno:2017}.  Furthermore you would still need to renormalize, and the implied subtraction process could easily vitiate the WEC for $\langle T^{ab}(\eta,\phi) \rangle$. }

Consider for instance the Casimir effect between parallel plates: The classical Maxwell Lagrangian certainly satisfies the WEC, but the renormalized electromagnetic energy density $\langle \rho\rangle$ is certainly negative~\cite{Visser:book, Zeta, Liberati:2001, Martin-Moruno:2013a, Martin-Moruno:2013b, twilight, Liberati:2000, Visser:2016, Martin-Moruno:2017}.

\section{Fixed-background curved-space: Simplified construction}

When investigating curved space QFT we find it most convenient to first start the discussion with a simple and direct construction for small $\epsilon$, subsequently extended to large $\epsilon$. We shall 
then present a generalized construction for large $\epsilon$, and ultimately circle back from the generalized large-$\epsilon$ construction to a generalized small-$\epsilon$ construction. We then introduce a tetrad version of the construction, which allows one to generalize the whole process.

\subsection{Small $\epsilon$}

Much of the previous flat-space discussion can be carried over to fixed-background curved-space QFT. While one can no longer Fourier transform to define the original momentum-space version of Feynman's $i\epsilon$ prescription, one can certainly define an almost-real $i\epsilon$-complexified (inverse) metric 
\begin{equation}
(g_\epsilon)^{ab} = (g_L)^{ab} -i\epsilon\,V^a\,V^b +\O(\epsilon^2).
\end{equation}
Then for the covariant metric
\begin{equation}
(g_\epsilon)_{ab} = (g_L)_{ab} +i\epsilon\,V_a\,V_b 
+\O(\epsilon^2).
\end{equation}


Here the vector field $V$ is now a possibly position-dependent unit 4-vector field, (a 4-velocity field), with respect to the Lorentzian metric $(g_L)_{ab}$, and $\epsilon(x)$ is allowed to be position-dependent but should at least be positive. While this is not the central point of the current article, this construction can now be used as the starting point for how to define a general notion of how to Wick rotate a spacetime metric~\cite{Visser:wick}.
See also~\cite{Barbero:1995,  Samuel:2015, Helleland:2015, Gray, Kothawala:2017, Singh:2020}, and related work~\cite{Girelli:2008,White:2008, Rajan:2016a}.
For instance, as long as one has a globally defined (real) co-tetrad $e^A{}_a$~\cite{Rajan:2016b}, then one could define
\begin{equation}
(g_\epsilon)_{ab} = (\eta_\epsilon)_{AB} \;\; e^A{}_a\; e^B{}_b,
\end{equation}
thereby forcing all of the complex structure into the tangent space metric $(\eta_\epsilon)_{AB}$.
We shall not (initially) take this specific route in the current article; but will ultimately develop and investigate a very closely related tetrad-based construction.

To analyze this slightly complex metric it is useful to note that
\begin{equation}
\det\left\{ (g_L)^{ab} (g_\epsilon)_{bc} \right\} 
= \det\left\{ \delta^a{}_c + i\epsilon \, V^a \, V_c\right\} = 1- i \epsilon +\O(\epsilon^2).
\end{equation}
That is
\begin{equation}
-\det\left\{ (g_\epsilon)_{ab} \right\} =
-\det\left\{ (g_L)_{ab} \right\}  (1- i \epsilon +\O(\epsilon^2) ).
\end{equation}
Thence
\begin{eqnarray}
\sqrt{-\det\left\{ (g_\epsilon)_{ab} \right\}} 
&=&
\sqrt{-\det\left\{ (g_L)_{ab} \right\}} \; \sqrt{1- i \epsilon +\O(\epsilon^2)}
\nonumber\\
&=& 
\sqrt{-\det\left\{ (g_L)_{ab} \right\}} \left\{1- {i \epsilon\over2} + \O(\epsilon^2)\right\}.
\end{eqnarray}
We can rewrite this more compactly as 
\begin{equation}
\label{E:det-small}
\sqrt{-g_\epsilon} 
=
\sqrt{-g_L} \; \sqrt{1- i \epsilon +\O(\epsilon^2)}
= 
\sqrt{-g_L} \; \left\{1- {i \epsilon\over2} + \O(\epsilon^2)\right\}.
\end{equation}
Thence we have
\begin{equation}
\sqrt{-g_\epsilon}\;  (g_\epsilon)^{ab} 
= \sqrt{-g_L} \left\{1- {i \epsilon\over2} + \O(\epsilon^2)\right\} \left[  (g_L)^{ab} -i\epsilon\,V^a\,V^b +\O(\epsilon^2) \right].
\end{equation}
We can expand this as 
\begin{equation}
\sqrt{-g_\epsilon} \;  (g_\epsilon)^{ab} 
= \sqrt{-g_L} 
\left[  (g_L)^{ab} -i\epsilon\,\left\{ V^a\,V^b +{1\over2} (g_L)^{ab}  \right\} +\O(\epsilon^2) \right].
\end{equation}
But if we now define
\begin{equation}
(g_E)^{ab} = (g_L)^{ab} + 2 V^a V^b,
\end{equation}
then $(g_E)^{ab}$ is a Euclidean signature metric, and furthermore
\begin{equation}
\det\left\{ (g_L)_{ab} (g_E)^{bc} \right\} = \det\left\{ \delta_a{}^c + 2 V_a V^b \right\} = -1.
\end{equation}
That is, we see that in this particular situation we have $\det\{(g_E)^{ab}\} = -\det\{(g_L)^{ab} \}$, or more briefly $g_E = - g_L$.
But then we can write
\begin{equation}
\sqrt{-g_\epsilon}  (g_\epsilon)^{ab} 
= \sqrt{-g_L} \, (g_L)^{ab} -i\,{\epsilon\over2}\, \sqrt{g_E} \, (g_E)^{ab} 
+\O(\epsilon^2). 
\end{equation}

From (\ref{E:det-small}) and the above  we have
\begin{equation}
\Im \left\{ \sqrt{-g_\epsilon}\right\} =
-{\epsilon\over2} \sqrt{-g_L} + \O(\epsilon^2);
\end{equation}
and
\begin{equation}
\Im \left\{ \sqrt{-g_\epsilon} \; (g_\epsilon)^{ab} \right\}
=
-{\epsilon\over2} \sqrt{g_E}\; (g_E)^{ab}  +\O(\epsilon^2);
\end{equation}
both of which are negative as long as $\epsilon$ is positive.

But these are simply small-$\epsilon$ versions of the Louko--Sorkin acceptability conditions~\cite{Louko:1995}:
\begin{equation}
\Im \{ \sqrt{-g_\epsilon}\}  <0; \qquad \qquad
\Im \left\{ \sqrt{-g_\epsilon} \; (g_\epsilon)^{ab}\right\} < 0.
\end{equation}
See particularly the footnote on page 193 of the Louko--Sorkin~\cite{Louko:1995} article.
These two conditions were explored by Louko and Sorkin~\cite{Louko:1995}, almost 25 years ago,  to guarantee the quantum stability of a (free) massive scalar field described by the action
\begin{equation}
S(g_\epsilon,\phi) =  {1\over2} \int \sqrt{-g_\epsilon} \; \left\{ - (g_\epsilon)^{ab} \partial_a \phi \; \partial_b \phi - m^2 \phi^2 \right\}.
\end{equation}
In the language of  Kontsevich and Segal~\cite{Segal:2021}, and Witten~\cite{Witten:2021} these are the 0-form and 1-form constraints. In $(3+1)\sim4$ dimensions there is also a 2-form constraint, which we shall discuss in due course.

It is perhaps worth noting that the the combination $\sqrt{-g_\epsilon} \; (g_\epsilon)^{ab}$ also shows up elsewhere in general relativity. Specifically, in defining the 
harmonic gauge $\Box\, x^a =0$, which is a coordinate gauge fixing condition equivalent to $\partial_a \{\sqrt{-g_\epsilon} \; (g_\epsilon)^{ab}\}=0$. This is the strong-field generalization of the weak-field Einstein--Hilbert--Fock--De~Donder gauge. 

Louko and Sorkin then partially generalized their construction~\cite{Louko:1995},  by defining the equivalent of 
\begin{equation}
(g_\epsilon)_{ab} = (g_L)_{ab} +i\epsilon\; \sum_{j} w_j \; (V^j)_a\,(V^j)_b.
\end{equation}
Here one is now performing a positive-weight ($w_j>0$) sum over timelike (or null) vector fields $(V^j)_a$. 
(See equation (5.1) of the Louko--Sorkin~\cite{Louko:1995} article.)
But can we generalize this construction even further?
That is our goal in the next subsection.

\subsection{Large $\epsilon$}

First let us verify that the complex metric $(g_\epsilon)_{ab} = (g_L)_{ab} +i\epsilon V_a\,V_b$ satisfies the two Louko--Sorkin constraints for arbitrarily large $\epsilon>0$, and then further generalize the construction. 
Note that, (adopting Riemann normal coordinates and going to the rest frame of the 4-velocity $V$), the eigenvalues of $(g_\epsilon)_{ab}$ are
\begin{equation}
\lambda \in \left\{ -1+i\epsilon,1,1,1 \right\}.
\end{equation}
Then $\sqrt{-g_\epsilon} = \sqrt{1-i\epsilon}$ and the eigenvalues of $\sqrt{-g_\epsilon} \;(g_\epsilon)^{ab}$ are:
\begin{equation}
\lambda \in \left\{ -{1\over\sqrt{1-i\epsilon}},\sqrt{1-i\epsilon},\sqrt{1-i\epsilon},\sqrt{1-i\epsilon} \right\}.
\end{equation}
The two Louko--Sorkin acceptability conditions are then
\begin{equation}
\Im (\sqrt{1-i\epsilon}) <0; \qquad
\Im\left( -{1\over\sqrt{1-i\epsilon}}\right) <0.
\end{equation}
But
\begin{equation}
\Im\left( -{1\over\sqrt{1-i\epsilon}}\right) = 
\Im\left( -{1\over\sqrt{1-i\epsilon}} \; {\sqrt{1+i\epsilon}\over\sqrt{1+i\epsilon}}\right) =
{\Im(-\sqrt{1+i\epsilon})\over\sqrt{1+\epsilon^2}} 
= {\Im(\sqrt{1-i\epsilon})\over\sqrt{1+\epsilon^2}}.
\end{equation}
So, for the specific metric $(g_\epsilon)_{ab} = (g_L)_{ab} +i\epsilon V_a\,V_b$, the two Louko--Sorkin acceptability conditions are actually degenerate, and amount to the \emph{same} assertion that $\epsilon>0$. This now works for arbitrarily large values of $\epsilon$.

\subsection{Rephrasing the construction in terms of tetrad formalism}

Let us now collect these ideas,  and recast them in terms of the tetrad formalism. Let $e_A{}^a(x)$ denote an arbitrary (not necessarily orthonormal) tetrad. The only things we need to insist on is that the contravariant vectors $e_A{}^a(x) = [e^a(x)]_A$ form a basis for the tangent space.
Thence in particular $\det(e_A{}^a(x))\neq 0$, and the co-tetrad $e^A{}_a(x)$ is simply the matrix inverse of $e_A{}^a(x)$, with the co-vectors $e^A{}_a(x) = [e_a]^A$ now forming a basis for the cotangent space. 

Now let $\theta(x)$ be any smooth function bounded by
$0< \theta(x) < \pi/2$. The variable $\theta(x)$ is effectively a nonlinear way of encoding the parameter $\epsilon(x)$ via the relation $\theta(x) = {1\over2} \arctan \epsilon(x)$; we shall soon see why this is worthwhile.
Define
\begin{equation}
(\eta_\theta)_{AB} = \diag\{ - \exp(-i2\theta(x)), 1,1,1\};
\end{equation}
and
\begin{equation}
(\eta_\theta)^{AB} = \diag\{ - \exp(+i2\theta(x)), 1,1,1\}.
\end{equation}

Construct the complex metric
\begin{eqnarray}
(g_\theta)_{ab} &=& (\eta_\theta)_{AB} \;e^i{}_A(x)\; e^B{}_b(x) 
\\ &=& 
- \exp[-i2\theta(x)] \; e^0{}_a(x) \; e^0{}_b(x)
+ \sum_{i=1}^3 e^i{}_a(x)\; e^i{}_b(x).
\end{eqnarray}
Then the inverse metric is
\begin{eqnarray}
(g_\theta)^{ab} &=& (\eta_\theta)^{AB} \;  e_A{}^a(x)\; e_B{}^b(x) 
\\ &=&
- \exp[+i2\theta(x)] \; e_0{}^a(x) \; e_0{}^b(x)
+ \sum_{i=1}^3 e_i{}^a(x)\; e_i{}^b(x).
\end{eqnarray}
Note that $\sqrt{-\det(g_\theta)} = \exp\left[-i\theta(x)\right]$ and thence 
\begin{eqnarray}
\sqrt{-\det(g_\theta)} \; (g_\theta)^{ab} &=& 
\exp\left[-i\theta(x)\right]\;
(\eta_\theta)^{AB} \;  e_A{}^a(x)\; e_B{}^b(x) 
\\ &=&
- \exp(+i\theta(x)) \; e_0{}^a(x) \; e_0{}^b(x)
+ \exp(-i\theta(x))  \sum_{i=1}^3 e_i{}^a(x)\; e_i{}^b(x).
\nonumber\\
&&
\end{eqnarray}
Thence, this rather general class of complex metrics manifestly satisfies the Louko--Sorkin (0-form and 1-form) acceptability conditions, and we shall soon see that it also satisfies the Louko--Sorkin, Kontsevich--Segal and Witten 2-form acceptability condition.
This is not the most general class of complex metrics satisfying the compatibility conditions, but it is a good basis for ultimately deriving the most general class of such complex metrics.

The corresponding Lorentzian and Euclidean metrics are given by
\begin{equation}
\sqrt{- g_L } \; (g_L)^{ab} =
\cos[\theta(x)] \left\{ - e_0{}^a(x) \; e_0{}^b(x) +
 \sum_{i=1}^3 e_i{}^a(x)\; e_i{}^b(x)\right\},
\end{equation}
and
\begin{equation}
\sqrt{g_E } \; (g_E)^{ab} =
\sin[\theta(x)] \left\{ + e_0{}^a(x) \; e_0{}^b(x) +
 \sum_{i=1}^3 e_i{}^a(x)\; e_i{}^b(x)\right\}.
\end{equation}
We shall ultimately seek to fully generalize this construction.

\section{Fixed-background curved-space: Generalized construction}

We shall now generalize the construction beyond simply using the 4-velocity vector $V^a$, to allow for a more complicated 
matrix-based  relation between $(g_L)_{ab}$ and $(g_E)_{ab}$. 
After dealing with the 0-form and 1-form acceptability conditions (the 2-dimensional Louko--Sorkin conditions), we then turn to the additional 2-form acceptability condition relevant in 4 dimensions.
\smallskip\null

\clearpage
\subsection{Large $\epsilon$}

Let us now construct the general solution to the two Louko--Sorkin acceptability conditions, allowing arbitrarily 
 ``large'' values of $\epsilon$; so we are no longer demanding that the metric is ``close'' to real. Take the Louko--Sorkin conditions as primary~\cite{Louko:1995}:
\begin{equation}
\label{E:primary}
\Im \{ \sqrt{-g_\epsilon}\}  <0; \qquad \qquad
\Im \left\{ \sqrt{-g_\epsilon} \; (g_\epsilon)^{ab}\right\} < 0.
\end{equation}

We note that $\left\{ \sqrt{-g_\epsilon} \; (g_\epsilon)^{ab}\right\}$ is by construction a tensor density of weight +1. Since the coordinates (and so the Jacobi matrices and determinants are real), its real and imaginary parts are both tensor densities of weight +1.  Thence the quantity $\Im \left\{ \sqrt{-g_\epsilon} \; (g_\epsilon)^{ab}\right\}$ is by construction a tensor density, and since it is known to  be a negative definite matrix, we can use it to \emph{define}  a Euclidean metric:
\begin{equation}
- \Im \left\{ \sqrt{-g_\epsilon} \; (g_\epsilon)^{ab}\right\}
= \epsilon \; \sqrt{g_E} \; (g_E)^{ab}.
\end{equation}
This definition implies an implicit and somewhat subtle non-linear construction of $(g_E)^{ab}$. 
That this definition is nevertheless effective can be argued as follows. 
The quantity $X^{ab} = -\Im \left\{ \sqrt{-g_\epsilon} \; (g_\epsilon)^{ab}\right\}$ is by construction a positive definite tensor density of weight +1. Therefore $\det[X^{ab}]$ is by construction a positive scalar density of weight $+4-2=2$. 
(All eigenvalues are positive and there are four of them.) 
Therefore $X^{ab} /\sqrt{\det X^{ab}}$ is a true tensor (not a tensor density), which must still be positive definite. One then defines
\begin{equation}
(g_E)^{ab} \propto {-X^{ab} \over \sqrt{\det X^{ab}}} =
{ \Im \left\{ \sqrt{-g_\epsilon}\; (g_\epsilon)^{ab}\right\} \over \sqrt{\det[ \Im \left\{ \sqrt{-g_\epsilon}\; (g_\epsilon)^{ab} \right\} ] } }
\end{equation}
Consistency then demands
\begin{equation}
\det(g_E) \propto \det\left[ \Im \left\{ \sqrt{-g_\epsilon}\; (g_\epsilon)^{ab} \right\}\right].
\end{equation}

The overall scale of $ (g_E)^{ab}$ is arbitrary, as it can always be absorbed into a redefinition of  $\epsilon$, and \emph{vice versa}. 
Similarly, we note that $\Re \left\{ \sqrt{-g_\epsilon} \; (g_\epsilon)^{ab}\right\}$ is also by construction a tensor density of weight +1. Since the whole point of the exercise was to start with a Lorentzian metric $(g_L)^{ab}$ and somehow complexify it, the physically appropriate choice is to demand:
\begin{equation}
\Re \left\{ \sqrt{-g_\epsilon} \; (g_\epsilon)^{ab}\right\}
=  \sqrt{-g_L} \; (g_L)^{ab}.
\end{equation}
Analogous to what we did for $(g_E)_{ab}$ let us define $Z^{ab} = \Re \left\{ \sqrt{-g_\epsilon} \; (g_\epsilon)^{ab}\right\}$, a tensor density of weight +1, whuch does not have to be positive definite. Then $\det[Z^{ab}]$ is a scalar density of weight +2.  Therefore $Z^{ab} /\sqrt{|\det Z^{ab}|}$ is a true tensor (not a tensor density). Then we can set
\begin{equation}
(g_L)^{ab} = {Z^{ab} \over \sqrt{|\det Z^{ab}|}}; \qquad
\det(g_L) = \det(Z^{ab}).
\end{equation}
(This sort of construction is well-known in the GR community, where one often deals with the tensor density $H^{ab} = \sqrt{-g} \; g^{ab}$ and then has to reconstruct the inverse metric $g^{ab}$ from the tensor density $H^{ab}$.)

Combining these observations we have one of our main results:
\begin{itemize}
\item 
The Louko--Sorkin matrix condition, (the 1-form condition), can now be written as:
\begin{equation}
\label{E:LS1}
\sqrt{-g_\epsilon} \; (g_\epsilon)^{ab}
=  \sqrt{-g_L} \; (g_L)^{ab} - i \epsilon \; \sqrt{g_E} \; (g_E)^{ab}.
\end{equation}
\item 
The Louko--Sorkin determinant condition, (the 0-form condition), can, from (\ref{E:primary}), be written as
\begin{equation}
\label{E:LS2}
\Im \left( \sqrt{-\det\left\{
 \sqrt{-g_L} \; (g_L)^{ab} - i \epsilon \; \sqrt{g_E} \; (g_E)^{ab}
\right\}}
\right)
< 0.
\end{equation}
\end{itemize}
This is the cleanest \emph{theoretical} way I have come up with for encoding the Louko--Sorkin 0-form and 1-form acceptability conditions.
However, it is sometimes worthwhile to re-express these ``symmetric'' results in a slightly more asymmetric, but computationally more useful, manner.

Taking the determinant of equation (\ref{E:LS1}) we have
\begin{equation}
- g_\epsilon = - g_L \; \det\left\{
\delta_a{}^c -  i \epsilon \; {\sqrt{g_E}\over\sqrt{-g_L}} \; (g_L)_{ab}(g_E)^{bc} 
\right\}.
\end{equation}
This suggests that one can make some progress by defining the rescaled positive definite tensor (not a tensor density)
\begin{equation}
h^{ab} = {\sqrt{g_E}\over\sqrt{-g_L}} \; {(g_E)^{ab}}.
\end{equation}
Then the Louko--Sorkin determinant condition (\ref{E:LS2}) 
can now be  rewritten as
\begin{equation}
\label{E:LS2c}
\Im \left( \sqrt{\det\left\{
\delta_a{}^c -  i \epsilon \;  (g_L)_{ab} \, h^{bc} 
\right\}}
\right)
< 0.
\end{equation}
In contrast the Louko--Sorkin matrix condition (\ref{E:LS1}) can now be written as
\begin{equation}
\label{E:LS1b}
\sqrt{-g_\epsilon} \; (g_\epsilon)^{ab}
=  \sqrt{-g_L} \; \left\{ (g_L)^{ab} - i \epsilon \; h^{ab} \right\}.
\end{equation}\enlargethispage{40pt}
If we try to invert the general formula for $\sqrt{-g_\epsilon} \; (g_\epsilon)^{ab}$, to explicitly extract $ (g_\epsilon)^{ab}$ as a function of $(g_L)^{ab} $ and $h^{ab}$, then the best we can do is this:
\begin{equation}
\label{E:master}
 (g_\epsilon)^{ab} = {
 (g_L)^{ab} - i \epsilon \; h^{ab}
 \over
 \sqrt{\det\left\{
\delta_c{}^e -  i \epsilon \; (g_L)_{cd}h^{de} 
\right\}
}
}.
\end{equation}
This, together with the determinant condition (\ref{E:LS2c}),  now completely characterizes the set of Luoko--Sorkin ``acceptable'' complex metrics in terms of the underlying Lorentzian metric $ (g_L)^{ab} $ and a positive-definite distortion tensor $h^{ab}$. 
This is the cleanest \emph{practical} way I have come up with for encoding the Louko--Sorkin acceptability conditions.

\subsection{Back to small $\epsilon$}
One place where we can make significant further progress with equation (\ref{E:master}) is in the small-$\epsilon$ limit.
In this limit we note that the determinant can be approximated as
\begin{equation}
 (g_\epsilon)^{ab} = {
 (g_L)^{ab} - i \epsilon \; h^{ab}
 \over
 \sqrt{
1 -  i \epsilon \; (g_L)_{cd} h^{cd} 
+ \O(\epsilon^2)
}
}.
\end{equation}
Thence
\begin{equation}
 (g_\epsilon)^{ab} = 
 \left\{ (g_L)^{ab} - i \epsilon \; h^{ab}\right\}
 \left\{ 1 +  i \epsilon{1\over2} \; \; (g_L)_{cd}h^{cd} 
+ \O(\epsilon^2)
\right\}.
\end{equation}
That is
\begin{equation}
 (g_\epsilon)^{ab} = 
 (g_L)^{ab} - i \epsilon \; \left\{ h^{ab}
 -{1\over2} \; [(g_L)_{cd} h^{cd}]   (g_L)_{ab} \right\}
+ \O(\epsilon^2).
\end{equation}
But the quantity 
\begin{equation}
\overline{h^{ab}}
= 
 h^{ab}
 -{1\over2} \; \{ (g_L)_{cd}h^{cd} \}  (g_L)_{ab} 
\end{equation}
is the familiar ``trace-reversal'' process, commonly encountered when perturbing the spacetime metric around some chosen background.
Specifically, we now have
\begin{equation}
 (g_\epsilon)^{ab} = 
 (g_L)^{ab} - i \epsilon \; \overline{h^{ab}}
+ \O(\epsilon^2),
\end{equation}
where $h^{ab}$ is an arbitrary positive definite tensor.

\enlargethispage{20pt}
Furthermore the Louko--Sorkin determinant condition now reduces to inspecting
\begin{eqnarray}
\Im \left\{ 
\sqrt{
1 -  i \epsilon  \; (g_L)_{cd} \, h^{cd}  + \O(\epsilon^2)
}  
\right\} 
&=& 
\Im \left\{ 
1 -  i \epsilon  {1\over 2} \; (g_L)_{cd} \,h^{cd} + \O(\epsilon^2)
\right\}
\nonumber\\
&=& - \epsilon \; {1\over 2} \; (g_L)_{cd} \,h^{cd} + \O(\epsilon^3).
\end{eqnarray}
So to verify the Louko--Sorkin determinant condition one need merely check whether or not the trace $(g_L)_{cd} \, h^{cd} $ is positive.

The two formulae
\begin{equation}
(g_\epsilon)^{ab} = 
 (g_L)^{ab} - i \epsilon \; \overline{h^{ab}};
\qquad\qquad
(g_L)_{cd} \, h^{cd} > 0;
\end{equation}
represent the cleanest \emph{practical} way I have come up with for encoding the small-$\epsilon$ Louko--Sorkin acceptability conditions. If these two formulae are satisfied then there will at the very least be some finite interval $\epsilon\in(0,\epsilon_*)$ over which the Louko--Sorkin acceptability conditions are satisfied. Sometimes this can be extended to the entire positive half line $\epsilon\in(0,\infty)$. 

\subsection{The 2-form constraint}

In $(3+1)\sim 4$ dimensions there is also an additional  2-form constraint. See for instance  the discussion in the footnote on page 193 of the of the Louko--Sorkin~\cite{Louko:1995} article, and in the 
Kontsevich and Segal~\cite{Segal:2021}, and Witten~\cite{Witten:2021} articles.

\subsubsection{Large $\epsilon$}

The key idea is to demand that the electromagnetic, (or more generally, the non-abelian gauge fields), satisfy: 
\begin{equation}
\Im\left\{ \sqrt{-g_\epsilon} \; (g_\epsilon)^{ac} \;(g_\epsilon)^{bd} \; F_{ab}\; F_{cd} \right\} <0.
\end{equation}
The field strengths $F_{ab}$ are taken to be real, so this is equivalent to demanding that the object
\begin{equation}
M^{[ab][cd]} = -\Im\left\{ \sqrt{-g_\epsilon} \; 
\left[ (g_\epsilon)^{ac} \;(g_\epsilon)^{bd} 
- (g_\epsilon)^{ad} \;(g_\epsilon)^{bc} \right]
\right\}
\end{equation}
be a positive definite matrix when acting on 2-forms.
This can be converted into a statement about the eigenvalues $\lambda_A$ of $(g_\epsilon)_{ab}$. Based on the footnote on page 193 of the of the Louko--Sorkin~\cite{Louko:1995} article, and with minor changes in notation and conventions as compared to references~\cite{Segal:2021,Witten:2021}, the 2-form acceptability condition can be cast as
\begin{equation}
\Im \left\{ \sqrt{-g_\epsilon}\over \lambda_A \; \lambda_B \right\} < 0; \qquad (A\neq B)\; \qquad \{A,B\} \in [0..3].
\end{equation}
For completeness, when cast in this form  the two primary Louko--Sorkin conditions, the 0-form and 1-form acceptability conditions, are:
\begin{equation}
\Im \left\{ \sqrt{-g_\epsilon}\right\}<0; 
\qquad \hbox{and} \qquad
\Im \left\{ \sqrt{-g_\epsilon}\over \lambda_A \right\} < 0. 
\end{equation}
While we have seen how to interpret the 0-form and 1-form conditions geometrically, a simple geometric interpretation for the 2-form condition is trickier. 

Recall that from the 1-form condition (\ref{E:LS1b}) we had
\begin{equation}
\label{E:LS1bb}
\sqrt{-g_\epsilon} \; (g_\epsilon)^{ab}
=  \sqrt{-g_L} \; \left\{ (g_L)^{ab} - i \epsilon \; h^{ab} \right\}.
\end{equation}

Let us now go to Riemann local coordinates for the Lorentzian metric $(g_L)_{ab}$, then locally we have
\begin{equation}
\label{E:LS1bbb}
\sqrt{-g_\epsilon} \; (g_\epsilon)^{ab}
=   \eta^{ab} - i \epsilon \; \hat h^{ab} .
\end{equation}

Here the coordinate transformed $\hat h^{ab}$ is still positive definite, 
and because it is positive definite, it must be of Hawking--Ellis type I, and so can be diagonalized via a local Lorentz transformation. 
(To check this one merely needs to verify that Hawking--Ellis types II, III, and IV are \emph{not} positive definite. See for instance~\cite{Hell, core}.) 

Thus we can without loss of generality assume $\hat h^{ab}$ is diagonal, with positive diagonal entries $h_A$, ($A\in\{0,i\})$.
Then in this coordinate system, with this choice of basis, we have
\begin{equation}
\label{E:master-222}
 \sqrt{-g_\epsilon} \;(g_\epsilon)^{ab} =
 \diag\{-1,1,1,1\} - i \epsilon \; \diag\{h_0,h_1,h_2,h_3\}.
 \end{equation}
 Note that this manifestly satisfies the 1-form acceptability condition. 
 Furthermore, taking the determinant
 \begin{equation}
g_\epsilon = -(1+i\epsilon h_0) \prod\nolimits_ {i=1}^3 (1-i\epsilon h_i).
\end{equation}
So
\begin{equation}
\sqrt{-g_\epsilon} = 
\sqrt{(1+i\epsilon h_0) \prod\nolimits_ {i=1}^3 (1-i\epsilon h_i) }.
\end{equation}
Then the 0-form acceptability condition reduces to the nonlinear condition
\begin{equation}
\label{E:0-form}
\Im \left\{ \sqrt{(1+i\epsilon h_0) \prod\nolimits_ {i=1}^3 (1-i\epsilon h_i) } \right\} < 0,
\end{equation}
which constrains the values of $h_0$ and the $h_i$.
If one wishes the 0-form acceptability condition to hold for all $\epsilon>0$ then in particular it must hold for small $\epsilon$ and so one must demand $h_0<\sum_{i=1}^3 h_i $. This is equivalent to demanding $\eta_{ab} \, h^{ab} >0$.

\enlargethispage{40pt}
Furthermore the eigenvalues of $(g_\epsilon)_{ab}$ are then
\begin{equation}
-{\sqrt{-g_\epsilon}\over1+i\epsilon h_0}; 
\qquad {\sqrt{-g_\epsilon}\over1-i\epsilon h_i};
\end{equation}
and the 2-form acceptability conditions deduce to
\begin{equation}
\Im\left\{- {(1+i\epsilon h_0)(1-i\epsilon h_j)\over \sqrt{-g_\epsilon}} \right\} <0;
\qquad
\Im\left\{{(1-i\epsilon h_j)(1-i\epsilon h_k)\over \sqrt{-g_\epsilon}}\right\} < 0.
\end{equation}
Here $j\neq k$. 

But then, given our explicit formula for $g_\epsilon$, this can be rewritten as 
\begin{equation}
\label{E:2-form}
\Im\left\{ -\sqrt{(1+i\epsilon h_0)(1-i\epsilon h_j)\over (1-i\epsilon h_k)(1-i\epsilon h_l)} \right\} <0;
\qquad
\Im\left\{\sqrt{(1-i\epsilon h_j)(1-i\epsilon h_k)\over 
(1+i\epsilon h_0)(1-i\epsilon h_l)}\right\} < 0,
\end{equation}
where the indices $j$, $k$, and $l$ must all be distinct.
Taking the complex conjugate of the first inequality, we  can also write this as
\begin{equation}
\label{E:2-formb}
\Im\left\{ \sqrt{(1-i\epsilon h_0)(1+i\epsilon h_j)\over (1+i\epsilon h_k)(1+i\epsilon h_l)} \right\} <0;
\qquad
\Im\left\{\sqrt{(1-i\epsilon h_j)(1-i\epsilon h_k)\over 
(1+i\epsilon h_0)(1-i\epsilon h_l)}\right\} < 0.
\end{equation}
Using the identity
\begin{equation}
{1\over 1\pm i \epsilon h_A} = {1\mp i \epsilon h_A\over 1 +\epsilon^2 h_A^2} \propto 1\mp i \epsilon h_A, 
\end{equation}
and rearranging some indices, the two conditions in (\ref{E:2-formb}) actually collapse to one nonlinear constraint:
\begin{equation}
\label{E:2-form-final}
\Im\left\{ \sqrt{(1-i\epsilon h_0)(1+i\epsilon h_j)(1-i\epsilon h_k)(1-i\epsilon h_l)} \right\} <0,
\end{equation}
where the indices $j$, $k$, and $l$ must all be distinct.
Overall, the compatibility conditions now reduce to the statement that $h_0$ and the $h_i$ are all positive, and are subject to the constraints (\ref{E:0-form}) and (\ref{E:2-form-final}).  
 
One obvious solution to these acceptability constraints is $h_0 = h_i = 1$ with $\epsilon >0$. 
That is $\hat h^{ab} = \delta^{ab}$, whence after trace reversal $\overline{ h^{ab}} = \diag\{2,0,0,0\}$. But this just corresponds to $ (g_\epsilon)_{ab} = \eta_{ab} - i \epsilon V_a V_b$, which is where we started the discussion --- however now we see that this object also satisfies the 2-form acceptability condition. 
There are also many other solutions to the acceptability conditions. 
For instance, in the symmetric case $h_i = h_{space} $, taking  $h_0 < 3 h_{space}$ with $\epsilon>0$ also
satisfies both of these constraints.

Subject to these nonlinear constraints on the $h_A$, where $A\in\{0,i\}$, for a suitable tetrad $e_A{}^a$ we now have $\hat h^{ab} = \sum_{A=0}^3 h_A \;e_A{}^a \; e_A{}^b$, which is manifestly positive definite. This quantity can then be inserted into (\ref{E:LS1bbb}), to recreate $(g_\epsilon)_{ab}$ in the Riemann normal coordinate system, and thus, via (\ref{E:LS1bb}), implicitly recreate  the metric $(g_\epsilon)_{ab}$ in an arbitrary coordinate system. 
Overall, at this stage of the analysis we have a somewhat tedious set of nonlinear constraints on the $h^{ab}$, which in principle will fully characterize the allowable complex metrics $(g_\epsilon)_{ab}$. 
(These nonlinear constraints simplify considerably in the small-$\epsilon$ limit, and furthermore, we shall then introduce a variant of the tetrad formalism to effectively linearize them in general.)

\clearpage
\subsubsection{Small $\epsilon$}
If we want the 2-form constraints to hold 
for all $\epsilon>0$, then in particular for small $\epsilon$ we must demand $h_0+\sum_i h_i > 2h_j$, in addition to the 0-form result that $h_0 <  \sum_i h_i$. That is, for $\epsilon\to0^+$ the 2-form and 0-form constraints reduce to
\begin{equation}
h_0+\sum_i h_i > 2h_j; \qquad \qquad h_0 <  \sum_i h_i.
\end{equation}
These two constraints can be combined into the single formula
\begin{equation}
h_A < {1\over2} \sum_B h_B.
\end{equation}

If we now define $\langle h \rangle = {1\over4} \sum_B h_B$, then between the positivity constraint and the 0-form and 2-form constraints we have
\begin{equation}
0 < h_A < 2 \langle h \rangle,
\end{equation}
or equivalently 
\begin{equation}
| h_A - \langle h \rangle | < \langle h \rangle.
\end{equation}
We  again have $\hat h^{ab} = \sum_{A=0}^3 h_A \;e_A{}^a \; e_A{}^b$.
Thence, for small $\epsilon$ the matrix $h_{ab}$ is not just positive definite, but once the 0-form and 2-form constraints are taken into account, has very tightly interrelated eigenvalues --- $h_{ab}$ cannot deviate too far from being some multiple $h_{ab} \sim \langle h \rangle \; \delta_{ab}$ of the identity matrix. This now guarantees that there will at the very least be some finite interval $\epsilon\in(0,\epsilon_*)$ over which the all of the acceptability conditions are satisfied. Sometimes, as in the examples given at the end of the previous subsection, this can be extended to the entire positive half line $\epsilon\in(0,\infty)$. 

\subsection{Rephrasing the construction in terms of the tetrad formalism}

Let us recast this generalized construction in terms of the tetrad formalism. As for the simplified construction, let $e_A{}^a(x)$ denote an arbitrary (not necessarily orthonormal) tetrad; the co-tetrad $e^A{}_a(x)$ is simply the matrix inverse of $e_A{}^a(x)$. 

Now let $\theta_A(x)$ be any four smooth functions bounded by
$0< \theta_A(x) < \pi/4$. The variables $\theta_A$ are effectively a nonlinear way of encoding the parameter $\epsilon$ via the relations $\theta_A = \arctan (\epsilon h_A)$, we shall soon see why this is worthwhile.
Define
\begin{equation}
\theta_{AB} = \diag\{ \theta_A \},
\end{equation}
and the corresponding trace-reversed quantities
\begin{equation}
\bar \theta_{AB} = \diag\{ \bar \theta_A \} = \theta_{AB}  -{1\over2} \, \eta_{AB} \; (\eta^{CD} \theta_{CD}).
\end{equation}
Then $\bar{\bar\theta}_{AB} = \theta_{AB}$, trace-reversal is an involution.

Explicitly
\begin{equation}
\bar\theta_0 = {1\over2} \left(\theta_0 + \sum_{i=1}^3 \theta_i \right);
\end{equation}
\begin{equation}
\bar\theta_i = {1\over2} \left(\theta_i +\theta_0- \sum_{j\neq i} \theta_i \right).
\end{equation}

Now define the matrix
\begin{equation}
(\eta_{\bar\theta})_{AB} = 
\diag\{ - \exp(-i2\bar\theta_0(x)); \;\;\exp(+i2\bar\theta_i(x))\};
\end{equation}
and its matrix inverse
\begin{equation}
(\eta_{\bar\theta})^{AB} = 
\diag\{ - \exp(+i2\bar\theta_0(x));\;\;\exp(-i2\bar\theta_i(x))\}.
\end{equation}

Construct the complex metric
\begin{eqnarray}
(g_\theta)_{ab} &=& (\eta_{\bar\theta})_{AB} \;e^A{}_a(x)\; e^B{}_b(x) 
\\ &=& 
- \exp(-i2\bar\theta_0(x)) \; e^0{}_a(x) \; e^0{}_b(x)
+ \sum_{i=1}^3 \exp(+i2\bar\theta_i(x))\; e^i{}_a(x)\; e^i{}_b(x).
\quad
\end{eqnarray}
Then the inverse metric is
\begin{eqnarray}
(g_\theta)^{ab} &=& (\eta_{\bar\theta})_{AB} \;  e_A{}^a(x)\; e_B{}^b(x) 
\\ &=&
- \exp(+i2\bar\theta_0(x)) \; e_0{}^a(x) \; e_0{}^b(x)
+ \sum_{i=1}^3 \exp(-i2\bar\theta_i(x)) \;e_i{}^a(x)\; e_i{}^b(x).
\quad
\end{eqnarray}
Note that
\begin{eqnarray}
\sqrt{-\det(g_\theta)}  
&=& |\det(e^A{}_a)| \; \sqrt{-\det(\eta_{\bar\theta})} 
\nonumber\\
&=& |\det(e)| \; \exp\left(-i\eta^{AB} \bar \theta_{AB}\right)
\nonumber\\
&=&  |\det(e)| \; \exp\left(+i\eta^{AB} \theta_{AB}\right)
\end{eqnarray}
and thence 
\begin{eqnarray}
\sqrt{-\det(g_\theta)} \;\;  (g_\theta)^{ab} &=& 
\exp\left(-i\eta^{CD} \bar \theta_{CD} \right)
(\eta_{\bar \theta})_{AB} \;  |\det(e)| \; e_A{}^a(x)\; e_B{}^b(x) 
\\ &=&
- \exp(-2i\theta_0(x)) \; e_0{}^a(x) \; |\det(e)| \;e_0{}^b(x)
\nonumber\\
&&
+  \sum_{i=1}^3 \exp(+2i\theta_i(x)) \; |\det(e)| \;e_i{}^a(x)\; e_i{}^b(x).
\label{E:final}
\end{eqnarray}
(Carefully note that due to a subtle combination of phases, it is the $\theta_A$, not the $\bar\theta_A$, that finally occur in the expression (\ref{E:final}) for the tensor density $\sqrt{-\det(g_\theta)} \; (g_\theta)^{ab}$.)

The corresponding Lorentzian and Euclidean metrics are given by
\begin{equation}
\sqrt{- g_L } \; (g_L)^{ab} = |\det(e)| \;
\left\{ - \cos[2\theta_0 (x)] e_0{}^a(x) \; e_0{}^b(x) +
 \sum_{i=1}^3 \cos[2\theta_i(x)] e_i{}^a(x)\; e_i{}^b(x)\right\},
\end{equation}
and
\begin{equation}
\sqrt{ g_E } \; (g_E)^{ab} =
|\det(e)| \;
\left\{ + \sin[2\theta_0(x)] e_0{}^a(x) \; e_0{}^b(x) +
 \sum_{i=1}^3 \sin[2\theta_i(x)] e_i{}^a(x)\; e_i{}^b(x)\right\}.
\end{equation}

We note that:
\begin{itemize}
\item 
This construction satisfies the 1-form acceptability condition
 if and only if we have $\theta_A \in (0,\pi/4)$.
\item
This construction satisfies the 0-form acceptability condition
 if and only if we have $\eta^{AB} \theta_{AB} >0$, implying $\theta_0 < \sum_{i=1}^3 \theta_i$. 
\item
Finally, the construction satisfies the 2-form acceptability condition  if and only if in terms of the eigenvalues $\lambda_A$ of $(g_\theta)_{ab}$ we have
\begin{equation}
\Im \left\{ \sqrt{-g_\theta}\over \lambda_A \; \lambda_B \right\} > 0; \qquad (A\neq B). 
\end{equation}
\end{itemize}
But in terms of the $\theta_A$ variables, suitably modifying the discussion of the previous subsection,  this now translates to the bound
\begin{equation}
\theta_A < {1\over2} \sum_B \theta_B.
\end{equation}
Between the positivity (1-form) and the above (0-form plus 2-form) acceptability conditions we have
\begin{equation}
0< \theta_A < {1\over2} \sum_B \theta_B < \pi/2.
\end{equation}
We note
\begin{equation}
\sum_B \theta_B =  \sum_B \bar \theta_B 
 - (-\bar\theta_0 +\sum_i \bar\theta_i) = 2\bar\theta_0.
\end{equation}
For the $\bar\theta_A$ this leads to the less symmetrical looking conditions\footnote{Note this is compatible with the first simplified example we considered,  where with current normalization one has $\theta_{AB} = \theta\; \delta_{AB}$, with $0<\theta<\pi/4$, while $\bar\theta_{AB} = \diag\{2\theta,0,0,0\}$. }
\begin{equation}
0< \bar\theta_0 = {1\over 2} \sum_B \theta_B <\pi/2;
\qquad\qquad
|\bar\theta_i| < \min\{ \bar\theta_0, \pi/4\}.
\qquad
 \end{equation}

This construction now yields the most general class of complex metrics satisfying the compatibility conditions. By adopting the tetrad formalism, we have managed to construct a complete and comprehensive classification of \emph{all} the complex metrics satisfying the (4-dimensional) compatibility conditions.

\section{Fluctuating spacetime geometries}

When one wants to include fluctuating spacetime geometries in the functional integral the situation becomes \emph{much} messier. Fundamentally one is interested in
\begin{equation}
Z_L = \int \D g_L \; \exp\left( +i S(g_L) \right).    
\end{equation}
But to make the functional integral more plausibly convergent it seems better to integrate over Euclidean geometries:
\begin{equation}
Z_E = \int \D g_E \; \exp\left(- S(g_E) \right).    
\end{equation}
For an overview of the standard point of view in
Euclidean quantum gravity see references~\cite{Centenary, 300}. 
For some early work see~\cite{Gibbons:1977}.
For rather recent developments see~\cite{Segal:2021, Witten:2021, Lehners:2021}. 
For slightly heterodox points of view see~\cite{Visser:baby, Hayward:1995}, and~\cite[page 69]{Visser:book}.
See also the ``causal dynamical triangulation'' programme~\cite{Ambjorn:2004, Ambjorn:2005a, Ambjorn:2005b, Ambjorn:2006, Sotiriou:2011}.  

The heterodoxy has to do with selecting \emph{a priori} constraints on the configuration space of Euclidian geometries to functionally integrate over --- the partition function seems to be much better behaved when one integrates only over those Euclidean geometries that are compatible with the existence of a Lorentzian signature metric~\cite{Hayward:1995,Visser:baby, Ambjorn:2004, Ambjorn:2005a, Ambjorn:2005b, Ambjorn:2006, Sotiriou:2011}.

Note that our central result for allowable complex metrics, either in the symmetric form
\begin{equation}
\label{E:master-symmetric}
\sqrt{-g_\epsilon}\, (g_\epsilon)^{ab} = 
\sqrt{-g_L} \,(g_L)^{ab} 
- i \epsilon \sqrt{g_E}\, (g_E)^{ab},
\end{equation}
or in the asymmetric form
\begin{equation}
\label{E:master-asymmetric}
 (g_\epsilon)^{ab} = {
 (g_L)^{ab} - i \epsilon \; h^{ab}
 \over
 \sqrt{\det\left\{
\delta_c{}^e -  i \epsilon \; (g_L)_{cd}h^{de} 
\right\}
}
},
\end{equation}
will \emph{automatically} force us to work only with those manifolds that are compatible with the existence of a Lorentzian signature metric. 

One may wish to add additional constraints on the configuration space  of Lorentzian metrics $(g_L)_{an}$. Stable causality? Global hyperbolicity?
Fixed topological structure?  Unimodular (more precisely, fixed modulus) gravity? There is a veritable multitude of possible choices, but any such choices should be constrained by the explicit versions of the Louko--Sorkin and Kontsevich--Segal--Witten  acceptability conditions derived above. 

\enlargethispage{20pt}
\section{Conclusions}\label{S:discussion}

We have developed several fully explicit versions of the Louko--Sorkin, Kontsevich--Segal, and Witten (0-form,1-form, and 2-form) acceptability conditions for complex metrics on spacetime. We have back-tracked the derivation of these acceptability conditions to the fundamental physics of Feynman's $i\epsilon$ prescription, which we used to justify the notion of an ``almost real'' spacetime metric. 

These ``almost real'' metrics can then be extended to more general settings, where the complex metric is explicitly given in terms of the underlying Lorentzian metric, (the physical metric), and a positive-definite auxiliary tensor, which can be viewed as being proportional to a Euclidean metric. 

We then provided a tetrad construction explicitly codifying the set of all acceptable complex metrics.
Finally we briefly discuss the implications for the functional integral over spacetime geometries, and in the appendix provide a specific example of how the acceptability conditions can fail, even for finite-action solutions to the classical field equations.

\appendix
\section[Example: Explicit physically unacceptable complex metric]
{\leftline{Example: Explicit physically unacceptable complex metric}}\label{A:unacceptable}

It is worthwhile to give a specific and fully explicit example of a complex metric that is \emph{not} ``acceptable'', and does \emph{not} satisfy the Louko--Sorkin conditions. 
Based on the construction given by Witten~\cite{Witten:2021}, consider the specific metric
\begin{equation}
\d s^2 = - \d t^2 + \left( 1+ i \;{\d\epsilon(r)\over \d r}\right)^2 \d r ^2 
+ (r+i \epsilon(r))^2 \d^2 \Omega. 
\end{equation}
Here $\epsilon(r)$ is an even function of $r$ defined on the entire real line, 
with $\epsilon(r)=0$ for $|r| > a$ and $\epsilon(r)>0$ for $|r| < a$. 
For all $r$ this geometry is Riemann flat,\footnote{Consider the coordinate $\tilde r = r + i\epsilon(r)$.}  so it solves the (complex) Einstein equations.
For  $|r| > a$ one simply has two portions of Minkowski space, with the usual $(-+++)$ signature. So this is a wormhole geometry --- it is however not a \emph{Lorentzian}  (traversable) wormhole geometry, but is instead a ``complexified'' wormhole geometry. 

At the throat ($r=0$) we have by the assumption of evenness enforced $\d\epsilon/\d r \to 0$ so the signature there is $(-+--)$, there are 3 ``time'' directions. In the regions $r\in(0,a)$ and $r\in(-a,0)$ the geometry is explicitly complex and the usual notion of signature makes little to no sense. So we should expect this geometry to be pathological.

Indeed we have 
\begin{equation}
\sqrt{-g_\epsilon} = \left( 1+ i {\d\epsilon(r)\over \d r}\right) (r+i \epsilon(r))^2 \sin\theta
= {1\over 3} {\d\over \d r} [(r+i \epsilon(r))^3 ] \sin\theta.
\end{equation}
Thence
\begin{equation}
\Im\{ \sqrt{-g_\epsilon} \} = \left\{ {\d\epsilon(r)\over \d r}(r^2-\epsilon(r)^2) +2r\epsilon(r)
\right\}
\sin\theta.
\end{equation}
But this is by construction an odd function of $r$. 
So the Louko--Sorkin determinant condition (the 0-form condition) must be violated \emph{somewhere} in the interval $(-a,a)$. If it is satisfied at $+r$ it will be violated at $-r$ and \emph{vice versa}. 

The second Louko--Sorkin condition (the matrix condition, the 1-form condition) now amounts to investigating the matrix
\begin{eqnarray}
\sqrt{-g_\epsilon} \; (g_\epsilon)^{ab} &=& 
\left( 1+ i {\d\epsilon(r)\over \d r}\right) (r+i \epsilon(r))^2 \sin\theta
\nonumber\\
&& \qquad \times
\left[ \begin{array} {cccc}
-1 &0 &0 &0\\
0 & \left( 1+ i {\d\epsilon(r)\over \d r}\right)^{-2}& 0 &0\\
0 &0& (r+i \epsilon(r))^{-2} & 0\\
0&0&0& {(r+i \epsilon(r))^{-2}\over \sin^2\theta} 
\end{array}
\right].
\end{eqnarray}

Concentrate on the $tt$ part of the tensor density:
\begin{equation}
\sqrt{-g_\epsilon} \; (g_\epsilon)^{tt} = -\sqrt{-g_\epsilon}.
\end{equation}
Thence
\begin{equation}
\Im\{ \sqrt{-g_\epsilon} \; (g_\epsilon)^{tt} \} = -\Im{\sqrt{-g_\epsilon} }.
\end{equation}
But this is incompatible with the first Louko--Sorkin condition. Indeed we have already seen that this quantity is odd. 
So the second Louko--Sorkin condition is also violated somewhere in the interval $(-a,a)$. 
 
 We mention in passing that in any ultra-static spacetime (block diagonal metric, with $g_{tt}=-1$) the two Louko--Sorkin conditions are incompatible with each other, so no calculation was actually needed. 

Furthermore, consider the angular parts of the tensor density. For example 
\begin{equation}
\sqrt{-g_\epsilon} \; (g_\epsilon)^{\theta\theta} = \left( 1+ i {\d\epsilon(r)\over \d r}\right) \sin\theta.
\end{equation}
Whence
\begin{equation}
\Im\{ \sqrt{-g_\epsilon} \; (g_\epsilon)^{\theta\theta}\} = {\d\epsilon(r)\over \d r}\;  \sin\theta.
\end{equation}
But this is again an odd function of $r$. 
So again the second Louko--Sorkin condition is violated \emph{somewhere} in the interval $(-a,a)$. 

Similarly 
\begin{equation}
\sqrt{-g_\epsilon} \; (g_\epsilon)^{\phi\phi} = {1+ i {\d\epsilon(r)\over \d r} \over \sin\theta}.
\end{equation}
Whence
\begin{equation}
\Im\{ \sqrt{-g_\epsilon} \; (g_\epsilon)^{\phi\phi}\} ={1\over \sin\theta} \; {\d\epsilon(r)\over \d r}.
\end{equation}
But this is again an odd function of $r$. 
So again the second Louko--Sorkin condition is violated \emph{somewhere} in the interval $(-a,a)$.  

Consequently curved-space QFT on this particular complex wormhole background will exhibit uncontrolled anti-damping in the functional integral, and so this background should be rejected as physically unacceptable. 
Despite the fact that this geometry is a zero-action classical solution of the Einstein equations, this geometry should \emph{not} be used as a saddle point in the functional integral.

\section*{Acknowledgements}

MV was directly supported by the Marsden Fund, \emph{via} a grant administered by the
Royal Society of New Zealand.

\clearpage


\begin{thebibliography}{99}
\newcommand{\arXiv}[1]{arXiv:~{\href{https://arxiv.org/abs/#1}{\color{blue}#1}}}

\bibitem{Bjorken-Drell}
J. D. Bjorken and S. D. Drell, {\sl Relativistic 
Quantum Fields}, \\
(McGraw--Hill, New York, 1965).

\bibitem{Lifshitz}
E.~M.~Lifshitz and L.~P.~Pitaevskii,
{\em Relativistix Quantum Theory: Part 2},\\
(Pergamon, Oxford, 1973).

\bibitem{Itzykson-Zuber}
C. Itzykson and J.-B. Zuber, {\sl Quantum Field Theory}, \\
(McGraw--Hill, New York, 1980).

\bibitem{Peskin}
Michael Peskin and Daniel Schroeder,
{\sl An introduction to Quantum Field Theory},
(Westview Press,  USA, 1995)

\bibitem{Srednicki}
Mark Srednicki, {\sl Quantum Field Theory}, (Cambridge, England, 2007).


\bibitem{Candelas:1977}
P. Candelas and D. J. Raine, 
``Feynman propagator in curved space-time'', \\
Phys. Rev. D 15 (1977) 1494. 
\doi{10.1103/PhysRevD.15.1494}


\bibitem{Ivashchuk:1987}
V.D. Ivashchuk, ``Regularization by $\epsilon$-metric: I'',\\
Izvestiya Akademii Nauk Moldavskoi SSR.
Ser. Fiziko-tekhnicheskih i matematicheskih nauk. No. 3, p. 8-17 (1987). 
\\{} [In Russian. English translation available in reference~\cite{Ivashchuk:1987-2019}.]

\bibitem{Ivashchuk:1988}
V.D. Ivashchuk. ``Regularization by $\epsilon$-metric: II. The limit $\epsilon=0^+$'',\\
Izvestiya Akademii Nauk Moldavskoi SSR.
Ser. Fiziko-tekhnicheskih i matematicheskih nauk. No. 1, p. 10-20 (1988). 
\\{} [In Russian. English translation available in reference~\cite{Ivashchuk:1988-2002}.]

\bibitem{Ivashchuk:1987-2019}
V.~D.~Ivashchuk,
``Regularization by $\epsilon$-metric'',
[\arXiv{1902.03152} [hep-th]].\\{}
[English translation of reference~\cite{Ivashchuk:1987}.]

\bibitem{Ivashchuk:1988-2002}
V.~D.~Ivashchuk,
``Regularization by $\epsilon$-metric. II. 
Limit $\epsilon=0^+$'',\\{}
[\arXiv{2002.10527} [hep-th]].\\{}
[English translation of reference~\cite{Ivashchuk:1988}.]


\bibitem{Ivashchuk:1997}
V.D. Ivashchuk, ``Wick rotation, regularization of propagators by a complex metric and multidimensional cosmology'', Grav. Cosmol. 3 (1997) 8-16, \arXiv{gr-qc/9705008}

\bibitem{Visser:wick}
M.~Visser,
``How to Wick rotate generic curved spacetime'', GRF essay, 1991.
[\arXiv{1702.05572} [gr-qc]].




\bibitem{Glimm-Jaffe}
J. Glimm and A. Jaffe, {\sl Quantum Physics: A
Functional Integral Point of View}, (Springer-Verlag, New York, 1987).

\bibitem{Louko:1995}
J.~Louko and R.~D.~Sorkin,
``Complex actions in two-dimensional topology change'',
Class. Quant. Grav. \textbf{14} (1997), 179-204
\doi{10.1088/0264-9381/14/1/018}
[\arXiv{gr-qc/9511023} [gr-qc]].


\bibitem{Segal:2021}
M.~Kontsevich and G.~Segal,
``Wick rotation and the positivity of energy in Quantum Field Theory'',
Quart. J. Math. Oxford Ser. \textbf{72} (2021) no.1-2, 673-699
\doi{10.1093/qmath/haab027}
[\arXiv{2105.10161} [hep-th]].


\bibitem{Witten:2021}
E.~Witten,
``A note on complex spacetime metrics'',
[\arXiv{2111.06514} [hep-th]].


\bibitem{MTW}
C.~W.~Misner, K.~S.~Thorne, and J.~A.~Weeler, 
\emph{Gravitation}, \\
(Freeman, San Francisco, 1973).
ISBN13: 978-0691177793.


\bibitem{Greensite:1992}
J.~Greensite,
``Stability and signature in quantum gravity'',\\
Proceedings of the 1st Iberian Meeting on Gravity (IMG-1, 1992), 297-300.\\
ISBN-13: 978-9810213695
ISBN-10: 9810213697
  


\bibitem{Greensite:1993}
  J.~Greensite,
  ``Dynamical origin of the Lorentzian signature of space-time'',\\
  Phys.\ Lett.\ B {\bf 300} (1993) 34
  \doi{10.1016/0370-2693(93)90744-3}
  [\arXiv{gr-qc/9210008}].

\bibitem{Carlini:1993}
  A.~Carlini and J.~Greensite,
  ``Why is space-time Lorentzian?'',
  Phys.\ Rev.\ D {\bf 49} (1994) 866
  \doi{10.1103/PhysRevD.49.866}
  [\arXiv{gr-qc/9308012}].

\bibitem{Greensite:1994}
  J.~Greensite,
  ``Quantum mechanics of space-time signature'',\\
  Acta Phys.\ Polon.\ B {\bf 25} (1994) 5.

\bibitem{Greensite:1995}
A.~Carlini and J.~Greensite,\\
 ``Square root actions, metric signature, and the path integral of quantum gravity'',\\
  Phys.\ Rev.\ D {\bf 52} (1995) 6947
  \doi{10.1103/PhysRevD.52.6947}
  [\arXiv{gr-qc/9502023}].

\bibitem{Kothawala:2018}
D.~Kothawala,
``Euclidean Action and the Einstein tensor,''
Phys. Rev. D \textbf{97} (2018) no.12, 124062
\doi{10.1103/PhysRevD.97.124062}
[\arXiv{1802.07055} [gr-qc]].


\bibitem{Martin-Moruno:2017}
P.~Mart\'in-Moruno and M.~Visser,
``Classical and semi-classical energy conditions'',
Fundam. Theor. Phys. \textbf{189} (2017), 193-213
\doi{10.1007/978-3-319-55182-1\_9}
[\arXiv{1702.05915} [gr-qc]].

\bibitem{Martin-Moruno:2013a}
P.~Mart\'\i{}n-Moruno and M.~Visser,
``Classical and quantum flux energy conditions for quantum vacuum states'',
Phys. Rev. D \textbf{88} (2013) no.6, 061701
\doi{10.1103/PhysRevD.88.061701}
[\arXiv{1305.1993} [gr-qc]].

\bibitem{Martin-Moruno:2013b}
P.~Mart\'in-Moruno and M.~Visser,\\
``Semiclassical energy conditions for quantum vacuum states'',\\
JHEP \textbf{09} (2013), 050
\doi{10.1007/JHEP09(2013)050}
[\arXiv{1306.2076} [gr-qc]].

\bibitem{twilight}
C.~Barcel\'o and M.~Visser,
``Twilight for the energy conditions?'',\\
Int. J. Mod. Phys. D \textbf{11} (2002), 1553-1560
\doi{10.1142/S0218271802002888}
[\arXiv{gr-qc/0205066} [gr-qc]].

\bibitem{Visser:book}
M.~Visser,
``Lorentzian wormholes: From Einstein to Hawking'',\\
(AIP Press, now Springer, New York, 1995).

\bibitem{Zeta}
S.~Blau, M.~Visser and A.~Wipf,
``Zeta functions and the Casimir energy'',\\
Nucl. Phys. B \textbf{310} (1988), 163
\doi{10.1016/0550-3213(88)90059-4}\\{}
[\arXiv{0906.2817} [hep-th]].

\bibitem{Liberati:2001}
S.~Liberati, S.~Sonego and M.~Visser,\\
``Faster than $c$ signals, special relativity, and causality'',\\
Annals Phys. \textbf{298} (2002), 167-185
\doi{10.1006/aphy.2002.6233}\\{}
[\arXiv{gr-qc/0107091} [gr-qc]].

\bibitem{Liberati:2000}
S.~Liberati, S.~Sonego and M.~Visser,
``Scharnhorst effect at oblique incidence'',
Phys. Rev. D \textbf{63} (2001), 085003
\doi{10.1103/PhysRevD.63.085003}
[\arXiv{quant-ph/0010055} [quant-ph]].

\bibitem{Visser:2016}
M.~Visser,
``Regularization versus Renormalization: Why are Casimir energy differences so often finite?'',
Particles \textbf{2} (2018) no.1, 14-31
\doi{10.3390/particles2010002}
[\arXiv{1601.01374} [quant-ph]].


\bibitem{Barbero:1995}
  J.~F.~Barbero G.,
  ``From Euclidean to Lorentzian general relativity: The real way'',
  Phys.\ Rev.\ D {\bf 54} (1996) 1492
  \doi{10.1103/PhysRevD.54.1492}
  [\arXiv{gr-qc/9605066}].
  
  \bibitem{Samuel:2015}
  J.~Samuel,\\
  ``Wick Rotation in the Tangent Space'',
  Class.\ Quant.\ Grav.\  {\bf 33} (2016)  015006
  \doi{10.1088/0264-9381/33/1/015006}
  [\arXiv{1510.07365} [gr-qc]].

\bibitem{Helleland:2015}
  C.~Helleland and S.~Hervik,\\
  ``A Wick-rotatable metric is purely electric'',
  J. Geom. Phys. \textbf{123} (2018), 424-429
\doi{10.1016/j.geomphys.2017.09.015}
\arXiv{1504.01244} [math-ph].

\bibitem{Gray}
Finnian Gray,\\
``Black hole radiation, greybody factors, and generalised Wick rotation'',\\
MSc thesis, 2016, Victoria University of Wellington. 
\blue{\sf  http://hdl.handle.net/10063/5148} 

\bibitem{Kothawala:2017}
D.~Kothawala,
``Action and Observer dependence in Euclidean quantum gravity'',
Class. Quant. Grav. \textbf{35} (2018) no.3, 03LT01
\doi{10.1088/1361-6382/aa9fdf}
[\arXiv{1705.02504} [gr-qc]].

\bibitem{Singh:2020}
R.~Singh and D.~Kothawala,
``Geometric aspects of covariant Wick rotation,''
[\arXiv{2010.01822} [gr-qc]].


\bibitem{Girelli:2008}
  F.~Girelli, S.~Liberati and L.~Sindoni,\\
  ``Emergence of Lorentzian signature and scalar gravity'',\\
  Phys.\ Rev.\ D {\bf 79} (2009) 044019
  \doi{10.1103/PhysRevD.79.044019}\\{}
  [\arXiv{0806.4239} [gr-qc]].

\bibitem{White:2008}
  A.~White, S.~Weinfurtner and M.~Visser,\\
  ``Signature change events: A Challenge for quantum gravity?'',\\
  Class.\ Quant.\ Grav.\  {\bf 27} (2010) 045007
  \doi{10.1088/0264-9381/27/4/045007}\\{}
  [\arXiv{0812.3744} [gr-qc]].

\bibitem{Rajan:2016a}
D.~Rajan and M.~Visser,
``Cartesian Kerr\textendash{}Schild variation on the Newman\textendash{}Janis trick'',
Int. J. Mod. Phys. D \textbf{26} (2017) no.14, 1750167
\doi{10.1142/S021827181750167X}
[\arXiv{1601.03532} [gr-qc]].

\bibitem{Rajan:2016b}
D.~Rajan and M.~Visser,
``Global properties of physically interesting Lorentzian spacetimes'',
Int. J. Mod. Phys. D \textbf{25} (2016) no.14, 1650106
\doi{10.1142/S0218271816501066}
[\arXiv{1601.03355} [gr-qc]].

\bibitem{Hell}
S.~W.~Hawking and G.~F.~R.~Ellis,
{\sl The Large Scale Structure of Space-Time},
(Cambridge University Press, Cambridge, 1973)
\doi{10.1017/CBO9780511524646}

\bibitem{core}
P.~Mart\'in-Moruno and M.~Visser,
``Essential core of the Hawking\textendash{}Ellis types'',
Class. Quant. Grav. \textbf{35} (2018) no.12, 125003
\doi{10.1088/1361-6382/aac147}
[\arXiv{1802.00865} [gr-qc]].


\bibitem{Centenary}
S. W. Hawking and W. Israel, {\sl General Relativity: An Einstein Centenary
Survey}, (Cambridge University Press, Cambridge, 1979),

\bibitem{300}
S. W. Hawking and W. Israel, {\sl 300 Years of Gravitation}, \\
(Cambridge University Press, Cambridge, 1987).

\bibitem{Gibbons:1977}
  G.~W.~Gibbons,
  ``The Einstein action of Riemannian metrics and its relation to quantum gravity and thermodynamics'',
  Phys.\ Lett.\ A {\bf 61} (1977) 3.
  \doi{10.1016/0375-9601(77)90244-4}

\bibitem{Lehners:2021}
J.~L.~Lehners,
``Allowable complex metrics in minisuperspace quantum cosmology'',
[\arXiv{2111.07816} [hep-th]].




 \bibitem{Visser:baby}
M.~Visser,
``Wormholes, baby universes and causality'',\\
Phys. Rev. D \textbf{41} (1990), 1116
\doi{10.1103/PhysRevD.41.1116}

\bibitem{Hayward:1995}
  S.~A.~Hayward,
  ``Complex lapse, complex action and path integrals'',\\
  Phys.\ Rev.\ D {\bf 53} (1996) 5664
  \doi{10.1103/PhysRevD.53.5664}
  [\arXiv{gr-qc/9511007}].

\bibitem{Ambjorn:2004}
J.~Ambjorn, J.~Jurkiewicz and R.~Loll,
``Emergence of a 4-D world from causal quantum gravity'',
Phys. Rev. Lett. \textbf{93} (2004), 131301
\doi{10.1103/PhysRevLett.93.131301}
[\arXiv{hep-th/0404156} [hep-th]].

 \bibitem{Ambjorn:2005a}
J.~Ambjorn, J.~Jurkiewicz and R.~Loll,
``Spectral dimension of the universe'',\\
Phys. Rev. Lett. \textbf{95} (2005), 171301
\doi{10.1103/PhysRevLett.95.171301}
[\arXiv{hep-th/0505113} [hep-th]].

\bibitem{Ambjorn:2005b}
J.~Ambjorn, J.~Jurkiewicz and R.~Loll,
``Reconstructing the universe'',\\
Phys. Rev. D \textbf{72} (2005), 064014
\doi{10.1103/PhysRevD.72.06401}
[\arXiv{hep-th/0505154} [hep-th]].

\bibitem{Ambjorn:2006} 
  J.~Ambjorn, J.~Jurkiewicz and R.~Loll,\\
  ``Quantum gravity, or the art of building spacetime'',\\
  In: D.~Oriti, (ed.) \emph{Approaches to quantum gravity}, 341-359
  [\arXiv{hep-th/0604212}].
  
  \bibitem{Sotiriou:2011}
T.~P.~Sotiriou, M.~Visser and S.~Weinfurtner,
``Spectral dimension as a probe of the ultraviolet continuum regime of causal dynamical triangulations'',
Phys. Rev. Lett. \textbf{107} (2011), 131303
\doi{10.1103/PhysRevLett.107.131303}
[\arXiv{1105.5646} [gr-qc]].
  
 \bigskip
 \hrule
 \hrule
 \hrule
 \bigskip 

\end{thebibliography}
\end{document}